\documentclass[10pt, a4paper, twoside, final, onecolumn]{icton}

\IEEEoverridecommandlockouts
\usepackage{amsmath}
\usepackage{epsfig}
\usepackage{array}
\usepackage{amsmath}
\usepackage{mathtools}
\usepackage{graphicx}
\usepackage{cite}
\usepackage{float}
\usepackage{placeins}
\usepackage{url}
\usepackage{enumerate}
\usepackage[centerlast]{caption}
\usepackage{pgfplots}
\usepackage{tikz}
\usepackage{amssymb}
\usepackage{xcolor}
\begin{document}

\title{Overview and Comparison of Nonlinear Interference Modelling Approaches in Ultra-Wideband Optical Transmission Systems}

\author{ \authorblockN{{\bfseries Daniel Semrau$^1$, Robert I. Killey$^1$, Polina Bayvel$^1$}}\\
\authorblockA{\itshape $^1$Optical Networks Group, Department of Electronic and Electrical Engineering, UCL (University College London), Torrington Place, London WC1E 7JE, United Kingdom\\
e-mail: uceedfs@ucl.ac.uk}\\
}

\maketitle
\thispagestyle{empty}



%

\begin{abstract}
\noindent The recent advances in modelling nonlinear interference of systems operating beyond the C-band are discussed. Estimation accuracy as well as computational complexity of current approaches are compared and addressed.
\end{abstract}

\begin {keywords}
optical fibre communication, ultra-wideband transmission, nonlinear interference, inter-channel stimulated Raman scattering 
\end{keywords}
\section{INTRODUCTION}
\noindent Lightwave transmission extending beyond the C-band, exploiting the entire bandwidth of installed fibres, is increasingly being considered as a cost-effective alternative to deploying new multi-core/-mode fibres. However, for such ultra-wide bandwidths, the non-instantaneous nature of the nonlinear fibre response becomes significant, giving rise to inter-channel stimulated Raman scattering (ISRS). ISRS effectively transfers power from high to lower frequencies within the same optical signal. Although the physics of the interaction between the Kerr effect and ISRS are well understood \cite{Blow,ISRS}, efficient low-complexity performance models for modern coherent systems were (until recently) not available. Such models are key for efficient link design, real-time optimisation and physical layer-aware networking. Recently, much attention has been drawn to the development of such models, particularly by extending the modulation format independent Gaussian noise (GN) model to account for ISRS \cite{SemrauOpEx, RobertsJLT, CantonoOFC, CantonoJLT, SemrauJLT, SemrauECOC,Semrauarxiv2, SemrauJLTcf, Poggioliniarxiv}. Extensions of the closed-form formalism for arbitrary, non-Gaussian modulation formats have been reported in \cite{Semrauarxiv}. Experimental demonstrations on the interaction between Kerr effect and ISRS followed up, in order to validate the theoretical predictions \cite{CantonoOFC, CantonoJLT,SaavedraSRS,MinoguchiSRS}. To date, a number of modelling approaches exist in integral as well as in closed-form, varying in accuracy and complexity.
\par 
\ 
In this paper, an overview of the recently proposed approaches of modelling nonlinear interference in ultra-wideband transmission is presented. The estimation accuracies are compared to split-step simulations and their mathematical and computational complexity are briefly discussed.
\section{MODELLING APPROACHES IN INTEGRAL FORM}
\noindent The GN model is a first-order solution with respect to the nonlinearity coefficient of the nonlinear Schr\"odinger equation assuming Gaussian modulation \cite{PoggioliniJLT}. The nonlinear interference (NLI) power is given as $P_\text{NLI}=\eta P_i^3$, with NLI coefficient $\eta$ and channel launch power $P_i$. To extend the conventional GN model to account for ISRS, it is assumed that the temporal gain dynamics of ISRS are negligible. This is motivated by the averaging of many independently modulated channels involved in the scattering process, smoothing the occurring ISRS gain in time. As a resulting simplification, ISRS can be modelled as a frequency- and distance-dependent signal power profile $\rho\left(z,f\right)$, which is obtained by solving the continuous-wave Raman equations \cite{ISRS}. The validity of this approach has recently been experimentally demonstrated in \cite{SaavedraSRS,MinoguchiSRS}. In general, approaches to include ISRS in the conventional GN model can be categorised into two groups. The first approach, termed the effective attenuation approach, is to approximate $\rho\left(z,f\right)$ with exponential decays, that have modified attenuation coefficients or effective lengths \cite{SemrauOpEx,CantonoOFC}. The second approach, termed the ISRS GN model, is to fully rederive the conventional GN model based on the exact signal power profile for higher accuracy but with increased complexity \cite{RobertsJLT, CantonoOFC, CantonoJLT,SemrauJLT,SemrauECOC}.
\subsection{The effective attenuation approach}
\label{sec:integral_effective_attenuation}
\noindent The first approach to extend the conventional GN model for ISRS, was the introduction of exponential decays with channel-dependent attenuation coefficients or effective lengths to model the effect of ISRS \cite{SemrauOpEx,CantonoOFC}. The advantage of the approach is that the expressions from the conventional, and moderately complex, GN model, can be used. In \cite{SemrauOpEx}, the use of channel-dependent, effective attenuation coefficients $\alpha_{\textnormal{eff},i}$ was proposed, that resemble the actual effective length present in the fibre span, as
\begin{align}
\begin{split}
\label{elength}
L_{\textnormal{eff},i} = \displaystyle\int_{0}^{L} \rho(\zeta,f_i)d\zeta = \frac{1-\textnormal{exp}(-\alpha_{\textnormal{eff},i} L)}{\alpha_{\textnormal{eff},i}},
\end{split}
\end{align}
where $f_i$, $L_{\text{eff},i}$ are  frequency and effective length of channel $i$ and $L$ is the span length. The approach is valid for lumped amplification in the weak ISRS regime, where the signal power profile is well approximated by an exponential decay. The advantage of the approach is its minor additional complexity with respect to the conventional GN model. The only additional complexity, to model ISRS, is to solve the Raman equations and then perform regression operations to obtain the coefficients $a_{\text{eff},i}$. Both can be carried out within seconds, yielding a low complexity prediction model. However, for large ISRS power transfers, the signal power profile is not accurately modelled by exponential decays and approximation errors are expected.
\subsection{The ISRS GN model in integral-form}
\noindent In order, to precisely account for arbitrary ISRS power transfers, as well as for distributed amplification, the GN model has been rederived to account for an arbitrary signal power profile \cite{RobertsJLT, CantonoOFC, CantonoJLT,SemrauJLT} as
\begin{equation}
\begin{split}
&\eta = \frac{16\gamma^2B_{i}}{27P_i^3}\int df_1\int df_2\ S_{\text{Tx}}(f_1,f_2,f_i)\left|\int_0^Ld\zeta\ \sqrt{\frac{\rho(\zeta,f_1)\rho(\zeta,f_2)\rho(\zeta,f_1+f_2-f_i)}{\rho(\zeta,f_i)}}e^{j\phi\left(f_1,f_2,f_i,\zeta\right)}\right|^2, 
\label{eq:general_G}
\end{split}
\end{equation}
where $\phi=-4\pi^2(f_1-f_i)(f_2-f_i)\left[\beta_2+\pi\beta_3(f_1+f_2)\right]\zeta$, $S_{\text{Tx}}=G_{\text{Tx}}(f_1)G_{\text{Tx}}(f_2)G_{\text{Tx}}(f_1+f_2-f_i)$ with $G_{\text{Tx}}(f)$ being the power spectral density (PSD) of the signal at the transmitter and $B_i$ is the channel bandwidth. Eq. \eqref{eq:general_G} can be used for multiple spans by interpreting $\rho\left(\zeta,f\right)$ as the signal power profile of the \textit{entire} link.
\par 
\ 
Eq. \eqref{eq:general_G} is exact to first-order for arbitrary power profiles, enabling the modelling of strong ISRS transfers and distributed amplification scenarios. The computational complexity is higher than the effective attenuation approach \ref{sec:integral_effective_attenuation}, due to an additional integration dimension over distance $\zeta$. Moreover, the Raman equations must be solved manually for every span along the transmission link and substituted in \eqref{eq:general_G}.
\par 
\ 
For lumped amplification and optical bandwidths of up to $15$ THz, the ISRS GN model can be written in analytical form, avoiding the necessity of solving the Raman equations, easing implementation and providing more insight into the underlying parameter dependencies. This is enabled by an approximate solution of the Raman equations \cite{ISRS}. This dramatically simplifies the description of ISRS which can then be parametrised by only a single parameter $C_r$, which is a linear regression of the Raman gain spectrum. The NLI coefficient can then be written as \cite{SemrauJLT,SemrauECOC}
\begin{equation}
\eta = \frac{16\gamma^2B_iG_1(f_i)}{27P_i^3}  \int df_1\int df_2\left|\sum_{k=1}^{n} \int_{0}^{L_k} d\zeta S_k\left(f_1,f_2,f_i\right) \frac{\hat{P}_{k}e^{-\alpha \zeta-\hat{P}_{k}C_{\text{r}} L_{\text{eff}}(f_1+f_2-f_i)}}{\int G_{k}(\nu)e^{-\hat{P}_{k}C_{\text{r}} L_{\text{eff}}\nu} d\nu}e^{j\phi\left(f_1,f_2,f_i,\tilde{L}_k+\zeta\right)}\right|^2
\label{eq:ISRSGNmodel}
\end{equation}
where $G_k$ and $\hat{P}_k$ are the signal PSD and total power launched into span $k$, $\tilde{L}_k$ is the accumulated distance of span $k$, $L_\text{eff}=\frac{1-\exp\left(-\alpha \zeta\right)}{\alpha}$ and $S_k=\sqrt[]{\frac{G_{k}(f_1)G_{k}(f_2)G_{k}(f_1+f_2-f_i)}{G_{k}(f_i)}}$. Eq. \eqref{eq:ISRSGNmodel} can be conveniently used to model the NLI in point-to-point transmission with different launch power distributions per span due to non-ideal gain equalisation as well as for complex ultra-wideband network scenarios \cite{SemrauECOC}. The computational complexity of \eqref{eq:ISRSGNmodel} is comparable to that of \eqref{eq:general_G}.
\begin{figure*}
   \centering
    \includegraphics[]{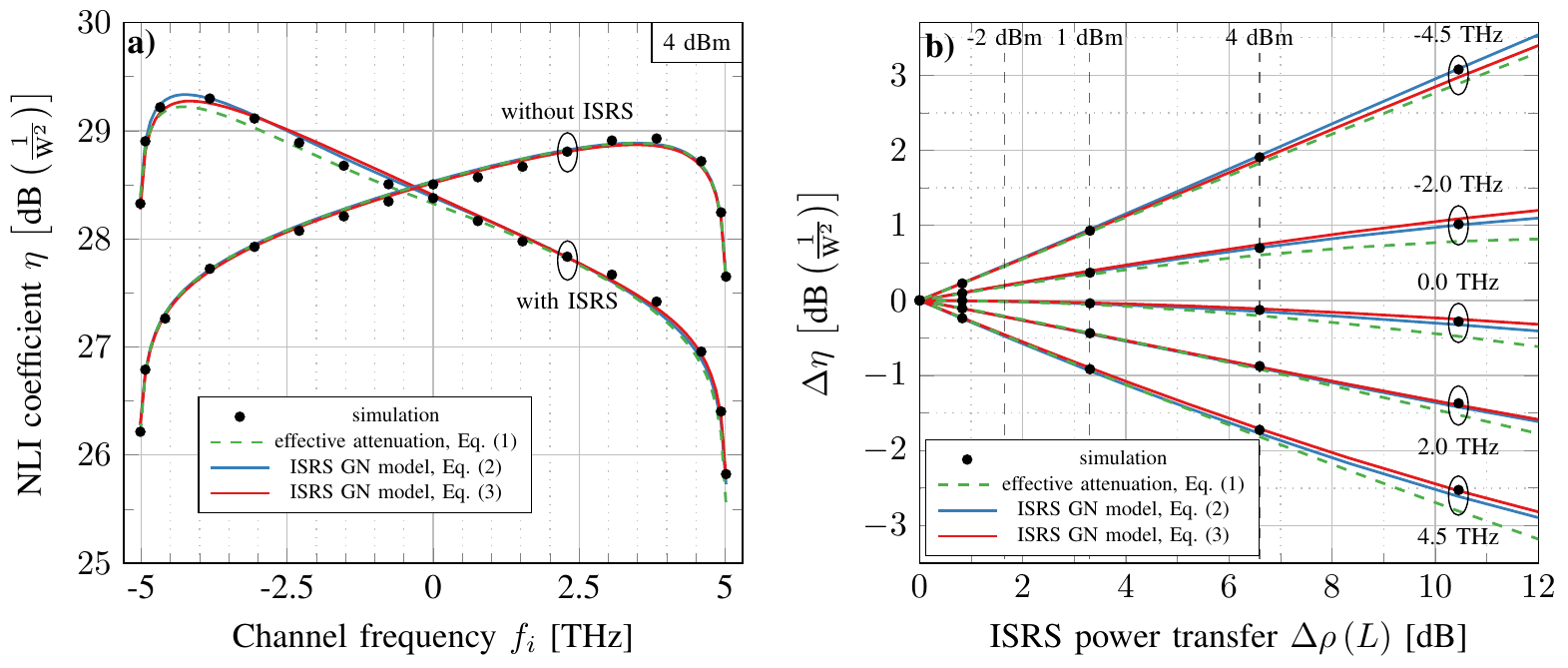}
\caption{NLI coefficient, and its deviation, after 3 spans obtained by integral approaches Eq. (1-3).}
\label{fig:integral}
\end{figure*}
\subsection{Comparison with split-step simulations}
\label{sec:split-step_setup}
\noindent To compare the accuracy of the modelling approaches, a split-step simulation was performed for $\text{119}\times\text{85}$~GBd Nyquist-spaced channels, occupying the entire C+L band (10.11~THz), centered at $\lambda_\text{ref}=1570$~nm. Transmission over a link comprising three spans of standard single mode fibre was considered, with parameters $\alpha=0.2 \frac{\text{dB}}{\text{km}}$, $D=18\frac{\text{ps}}{\text{nm}\cdot \text{km}}$, $S=0.067\frac{\text{ps}}{\text{nm}^2\cdot\text{km}}$, $\gamma=1.2\frac{\text{1}}{\text{W}\cdot\text{km}}$, $A_\text{eff}=81.8\mu\text{m}^2$ and $C_r=0.0236\frac{1}{\text{W}\cdot\text{km}\cdot\text{THz}}$. ISRS was implemented by a frequency-dependent loss at each simulation step, according to the power profile $\rho\left(f,\zeta\right)$ and ideal gain equalisation was performed. A sequence length of $2^{17}$ symbols was considered and four data realisations were averaged to increase simulation accuracy. In order to accurately benchmark the proposed models, Gaussian symbols were used for transmission. The NLI coefficient as a function of channel frequency and as a function of ISRS power transfer is shown in Fig. 1. The ISRS power transfer is defined as the sum of the ISRS gain/loss in decibel of the most outer WDM channels. The ISRS GN model in semi-analytical form \eqref{eq:general_G} and in analytical form \eqref{eq:ISRSGNmodel} have a negligible error with respect to the split-step simulations. The effective attenuation approach is in very good agreement with simulations despite the exponential decay approximation; with a maximum mismatch of 0.1 dB for 4~dBm launch power. However, this mismatch increases for increasing ISRS power transfers as ISRS does not strictly resemble an exponential decay.
\par 
\ 
In conclusion, the NLI estimates of all integral approaches are similar within 0.1 dB for powers up to 4 dBm per channel. However, for stronger ISRS power transfers and higher required accuracy, the use of the ISRS GN model is recommended at the expense of higher computational complexity.
\section{MODELLING APPROACHES IN CLOSED-FORM}
\noindent Although, the integral approaches above estimate the NLI with very good accuracy, they are not directly suitable for real-time applications and network optimisation where vast numbers of light paths must be evaluated. For such scenarios, approximations in closed-form, that yield results within picoseconds, have been proposed for the effective attenuation approach, as well as for the ISRS GN model \cite{SemrauOpEx,Semrauarxiv2,SemrauJLTcf,Poggioliniarxiv,Semrauarxiv}.
\subsection{The effective attenuation approach}
\noindent A semi-analytical closed-form approach was proposed in \cite{SemrauOpEx}. The approach is based on a closed-form solution of the conventional GN model extended by the effective attenuation as in Sec. \ref{sec:integral_effective_attenuation}. The NLI coefficient is approximated by \cite{SemrauOpEx}
\begin{align}
\begin{split}
\eta &\approx \frac{8}{27}\gamma^2 \frac{n^{1+\epsilon} \alpha_{\textnormal{eff},i}L_{\textnormal{eff},i}^2}{\pi |\beta_{2,i}| B_{i}^2}\textnormal{asinh}\left( \frac{0.5\pi^2 |\beta_{2,i}| B_{\textnormal{tot}}^2 }{  \alpha_{\textnormal{eff},i}}\right),
\label{asd}
\end{split}
\end{align}
with $\beta_{2,i}$ being the GVD parameter of channel $i$. Eq. \eqref{asd} is an approximate solution for the NLI of the central channel and approximation errors at the outer WDM channels are expected. The approach is semi-analytically, as the Raman equations have to be solved and regression functions have to be executed to obtain $\alpha_{\text{eff},i}$.
\subsection{The ISRS GN model in closed-form}
\noindent The first fully analytical closed-form approximation of the ISRS GN model was derived in \cite{Semrauarxiv2,SemrauJLTcf}. The formula accounts for arbitrary launch power distributions, channel configurations, ISRS and wavelength dependent attenuation and dispersion. The approximation relies on a first-order description of ISRS, introducing approximation errors for large ISRS power transfers, and is given by \cite{SemrauJLTcf}
\begin{equation}
\begin{split}
\eta &\approx \frac{4}{9}\frac{\gamma^2}{B^2_i} \frac{\pi n^{1+\epsilon} }{\phi_{i}\bar{\alpha}\left(2\alpha+\bar{\alpha}\right)}\cdot\left[\frac{T_i-\alpha^2}{a}\text{asinh}\left(\frac{\phi_{i}B_i^2}{\pi a}\right)+\frac{A^2-T_i}{A}\text{asinh}\left(\frac{\phi_{i}B_i^2}{\pi A}\right)\right]\\
&+\frac{32}{27}\sum_{k=1,k\neq i}^{N_\mathrm{ch}} \left(\frac{P_k}{P_i}\right)^2\frac{\gamma^2}{B_k}\frac{n}{\phi_{i,k}\bar{\alpha}\left(2\alpha+\bar{\alpha}\right)}\left[\frac{T_k-\alpha^2}{\alpha}\mathrm{atan}\left(\frac{\phi_{i,k}B_i}{\alpha}\right) +\frac{A^2-T_k}{A}\ \mathrm{atan}\left(\frac{\phi_{i,k}B_i}{A}\right)\right],
\label{eq:full_eta_cf}
\end{split}
\end{equation}
with $\phi_i=\frac{3}{2}\pi^2\left(\beta_2+2\pi\beta_3f_i\right)$, $T_i = \left(\alpha+\bar{\alpha}-P_{\text{tot}}C_{\text{r}}f_i\right)^2$ and $\phi_{i,k}=2\pi^2\left(f_k-f_i\right)\left[\beta_2+\pi\beta_3\left(f_i+f_k\right)\right]$ and $A=\alpha+\bar{\alpha}$. A formula based on the same assumptions with a similar result was presented in \cite{Poggioliniarxiv}. Eq. \eqref{eq:full_eta_cf} enables real-time NLI estimation in point-to-point links and mesh optical networks. Very recently, \eqref{eq:full_eta_cf} has been extended for arbitrary modulation formats \cite{Semrauarxiv}.
\subsection{Comparison with split-step simulations}
\noindent The NLI coefficient as a function of channel frequency and as a function of ISRS power transfer is shown in Fig. 2. While Eq. \eqref{asd} shows some approximation errors towards the outer WDM channels, it is reasonably accurate in calculating the change of the NLI coefficient for ISRS power transfers of up to 3~dB, with a maximum error of 0.2~dB compared to simulations. The ISRS GN model in closed-form \eqref{eq:full_eta_cf} shows very good agreement to simulations for all channels, with an average deviation of 0.1~dB at 4~dBm launch power. The maximum deviation of \eqref{eq:full_eta_cf} compared to the result in \cite{Poggioliniarxiv} was negligible ($<$0.1~dB) over the entire parameter space studied.
\par 
\ 
Although, \eqref{asd} offers some initial conclusions on the impact of ISRS on the NLI, the recently derived ISRS GN model in closed-form \eqref{eq:full_eta_cf} is superior in computational complexity as well as in accuracy. Therefore, we recommend the use of \eqref{eq:full_eta_cf} for the real-time modelling of ultra-wideband transmission systems.
\begin{figure*}
   \centering
    \includegraphics[]{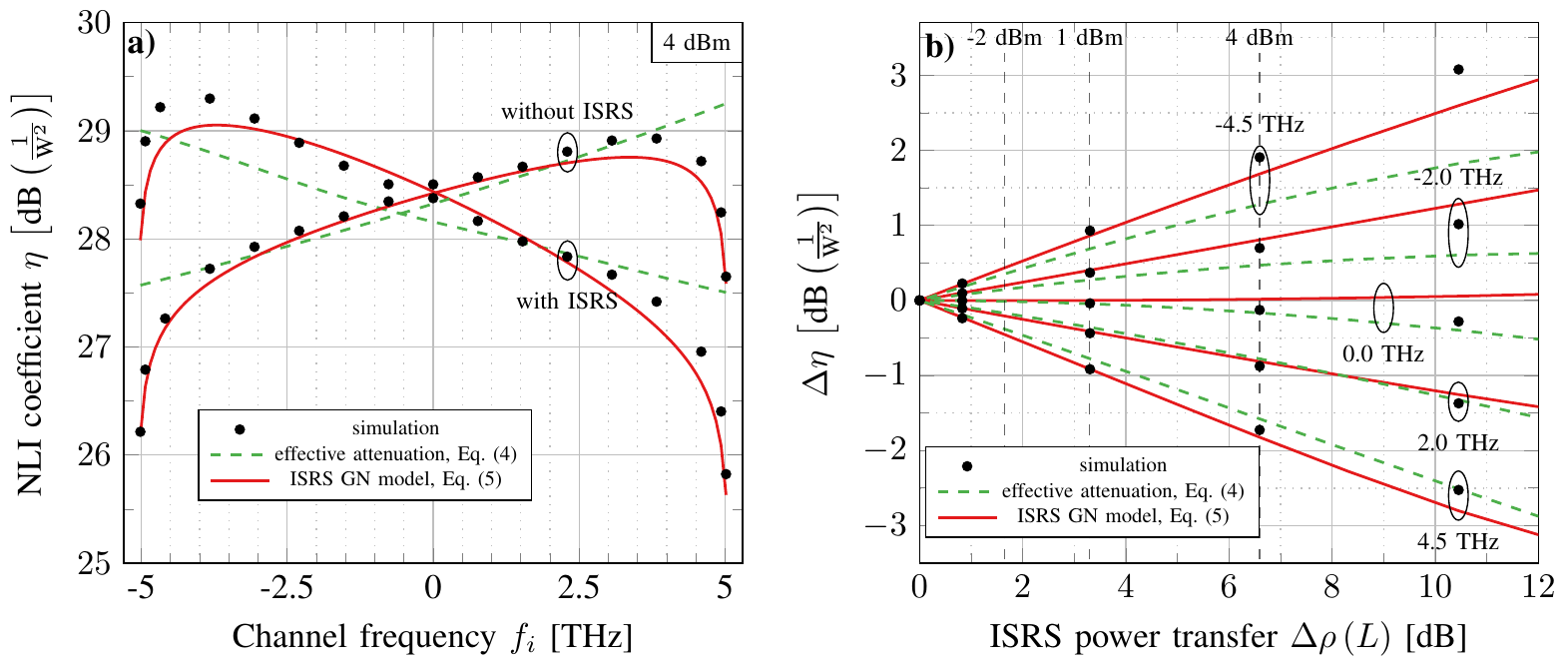}
\caption{NLI coefficient, and its deviation, after 3 spans obtained by closed-form approaches Eq. (4-5).}
\label{fig:cf}
\end{figure*}
\section{CONCLUSIONS}
\noindent The recently proposed approaches to account for inter-channel stimulated Raman scattering in nonlinear interference modelling are reviewed, in particular the effective attenuation approach and the ISRS GN model; both in integral and in closed-form. When the estimation accuracy is the priority, the ISRS GN model in integral form or the less complex effective attenuation approach should be considered. For time sensitive applications, the ISRS GN model in closed-form is recommended for almost instantaneous, but yet accurate, NLI estimation.
\section*{\small ACKNOWLEDGMENTS}
\noindent \small Financial support is from UK EPSRC DTG PhD studentship to D.~Semrau and EPSRC TRANSNET Programme Grant and is gratefully acknowledged.

\thispagestyle{empty}

\thispagestyle{empty}
\end{document}